\documentclass[preprint,showpacs,preprintnumbers,amsmath,amssymb]{revtex4-1}         %
\usepackage{graphicx,epsfig}% Include figure files
\usepackage{dcolumn}% Align table columns on decimal point
\usepackage{bm}% bold math
\newcommand{\be}{\begin{equation}}
\newcommand{\ee}{\end{equation}}
\newcommand{\bse}{\begin{subequations}}
\newcommand{\ese}{\end{subequations}}
\newcommand{\bary}{\begin{eqnarray}}
\newcommand{\eary}{\end{eqnarray}}
\newcommand{\bwt}{\begin{widetext}}
\newcommand{\ewt}{\end{widetext}}
\begin{document}

\title{Some possible sources of IceCube TeV-PeV neutrino events}
\author{Sarira Sahu,  Luis Salvador Miranda}
\affiliation{
Instituto de Ciencias Nucleares, Universidad Nacional Aut\'onoma de M\'exico,
Circuito Exterior, C.U., A. Postal 70-543, 04510 Mexico DF, Mexico}

%\date{\today} % It is always \today, today,

\begin{abstract}

The IceCube Collaboration has observed 37 neutrino events in the
energy range $30\, TeV\leq E_{\nu} \leq 2$ PeV
and the sources of these neutrinos are unknown. Here we have shown
that positions of 12 high energy blazars and the position of the FR-I
galaxy  Centaurus A, coincide within the error circles of ten IceCube
events, the later being in the error circle of the highest energy
event so far observed by IceCube. Two of the above blazars 
are simultaneously within the error circles of the Telescope Array
hotspot and one IceCube event. We found that the blazar
H2356-309 is within the error circles of three IceCube events.
We propose that  photohadronic interaction of the Fermi accelerated high energy protons
with the synchrotron/SSC background photons in the nuclear
region of these high energy blazars and AGN are probably responsible
for some of the observed IceCube events. 

\end{abstract}
\keywords{Neutrino \and HBL \and IceCube}
%\pacs{95.85.Ry, 98.70.Sa, 14.60.Lm } % PACS, the Physics and Astronomy
\maketitle

\section{Introduction}
\label{intro}

 In November 2012, the IceCube Collaboration announced  the detection of two
showerlike events with energies slightly above 1 PeV  by analyzing the
data taken during May 2010 - May 2012\cite{Aartsen:2013bka}.
%These data were taken using the detectors from 79 strings and completed 86 strings 
%with a total combined live time of 662 days. 
A follow-up analysis
of the same data published in November 2013, revealed additional 26 events in the energy range
$\sim$ 30 TeV - 250 TeV\cite{Aartsen:2013jdh}. Reconstruction of these events shows that 21
events are showerlike, mostly caused by $\nu_e$ and $\nu_{\tau}$ and 7 are muon track events. 
These 28 events have flavors, directions and energies
inconsistent with those expected from the atmospheric muon and
neutrino backgrounds and probably this is the first indication 
of extraterrestrial origin of high energy neutrinos. The track events
have uncertainty of order one degree  in their arrival directions and
the angular resolution for 21 shower events is poor, ranging from$\sim 10^{\circ}$
to $\sim 50^{\circ}$. The  IceCube  analysis  ruled out any spatial clustering of the events. The third
year (2012-2013) data analysis  revealed
additional 9 events of which two are track events and rest
are shower events \cite{Aartsen:2014gkd}. The event 35 is the most energetic one so far  observed.
In the full 988-day data, the muon background is expected to be
$8.4\pm4.2$ and the atmospheric neutrino is $6.6^{+5.9}_{-1.6}$. 
Five events 
%3, 8, 18, 28 and 32
are down going muons  and are consistent with the expected 
background muon events. This shows that the IceCube events are
predominantly shower events. 
%As it is anticipated, the atmospheric 
%neutrinos up to TeV energies exhibit a muon flavor dominance but the
%IceCube result contradicts this and is consistent with equal fluxes of
%all the three flavors. 
For a $E^{-2}_{\nu}$ spectrum the best fit diffuse flux obtained by
IceCube per flavor is $F_{\nu}=(0.95\pm 0.3)\times 10^{-8}
\,GeV\,cm^{-2}\,s^{-1}\,sr^{-1}$ which is consistent with the
Waxman-Bahcall bound\cite{Waxman:1998yy}.
Observation of these neutrinos  triggered
a lot of excitement to understand their origin and production mechanism.
While interpreting  these events in terms of astrophysical models seems
challenging, several possible galactic and extra galactic sources have been discussed 
which includes, Galactic
center\cite{Razzaque:2013uoa},
$\gamma$-ray bursts (GRBs)\cite{Murase:2013ffa}, active
galactic nuclei (AGN)\cite{Murase:2014foa}, high energy peaked blazars
(HBLs)\cite{ Padovani:2014bha, Krauss:2014tna},
starburst galaxies\cite{Murase:2013rfa} etc. 
In Ref.\cite{ Padovani:2014bha} many positional correlations of BL Lac
objects and galactic pulsar wind nebulae with the IceCube events are
shown. 
It is also  very natural to expect that these neutrinos
might come from diverse sources having different production mechanisms
and different power-law and this information can probably be extracted from the
directionality of the observed neutrino events. 
The largest
concentration of 7 events are around the Galactic center and also clustering of the events could be associated
to the Norma arm of the Galaxy\cite{Neronov:2013lza}. 
As the statistics is too sparse,
it is premature to draw any conclusion regarding the galactic origin
of these  events. 
There are also nonstandard physics interpretations
of these events\cite{Chen:2013dza,Anchordoqui:2013dnh}.

We found coincidence of 12 TeV emitting HBL positions and the FR-I galaxy
Centaurus A (Cen A) within the error circles of
10 IceCube events from the online catalog for TeV astronomy (TeVCat)\cite{TeVCat}.
cat. Due to the observed multi-TeV emission, these
objects are long believed to be sources of ultra high energy
cosmic rays (UHECRs). 
Few years ago, Pierre Auger (PA) collaboration reported two UHECR
events within $3.1^{\circ}$ around Cen A. 
%From the flaring of the blazars, neutrino flux is also estimated.
%Interaction of the UHECRs with the background
%photons and/or surrounding medium can produce neutrinos which can
%probably be observed by IceCube.
Therefore, in this work we focus our analysis on these candidate sources to find out
how the IceCube events with the desired energies can be produced
through photohadronic interaction within the core region of the
emanating jets. 

\section{Hadronic Model}  
\label{sec:1}

In the framework of the unification scheme of AGN, blazars and radio 
galaxies\cite{Ghisellini:1998it}, all are
intrinsically the same objects, viewed at different angles with respect
to the jet axis. The blazars have jets
pointing towards us. The double-peak spectral energy distribution (SED)
structure is common to all these objects. 
%The SED is explained through the leptonic model in which the emitting region in the jet is a blob with comoving
%radius $R'_b$ and moving with a Lorentz factor $\Gamma$ and a
%Doppler factor $\delta$.
%$\delta\simeq \Gamma^{-1} (1-\cos\theta_{ab})^{-1}$. 
%The first low-energy peak in radio to optical wavelength is the 
%synchrotron radiation from a population of relativistic electrons, 
%while the second high-energy peak in the  X-rays to very high energy
%(VHE) gamma-rays range is from the Compton scattering of the high
%energy population of electrons with the seed synchrotron photons in the jet. 
This model is successful
in explaining the multi-wavelength emission from BL Lac objects and FR-I
galaxies\cite{Tagliaferri:2008qk,Gutierrez:2006ak,Falcone:2010fk,Roustazadeh:2011zz,Blazejowski:2005ih}.
However, multi-TeV emission during flaring and non-flaring events from
these objects are difficult to reconcile.
Also  the most important challenge for the leptonic model
is to explain the orphan flaring from the blazars 1ES1959+650 and
Mrk 421. So variant of the hadronic models or the lepto-hadronic models are proposed to
explain these multi-TeV emissions.

The AGNs are efficient accelerators of
particles through shock or diffusive Fermi acceleration processes
with a power-law spectrum given as ${dN}/{dE} \propto E^{-\kappa}$,
with the power index $\kappa \geq 2$\cite{Dermer:1993cz}.
Protons can
reach ultra high energy (UHE) through the above acceleration
mechanisms. Fractions of these particles escaping from the source
can constitute the UHECRs arriving on Earth.
These objects also produce  high 
energy $\gamma$-rays and neutrinos through $pp$ and/or $p\gamma$ interactions 
\cite{Kachelriess:2008qx}.
%The core region of the jet (the blob) in the blazars and the FR-I galaxies are
%close to the supermassive black hole where seed photon density is
%high. So, if the protons in the region are accelerated enough, can
%undergo deep inelastic collision with the seed photons. On the other hand
%in the $pp$ process, the UHECR will collide with the low density hydrogen
%cloud far away from the core region which shows that in blazars
%(particularly in HBLs) and
%FR-I galaxies high energy $\gamma$-rays and neutrinos are produced
%more efficiently in $p\gamma$ process than in the $pp$ process.
The multi-TeV flaring events in 
blazars can be well explained by invoking hadronic model through $p\gamma$
interaction\cite{Sahu:2013cja,Sahu:2015tua,Sahu:2013ixa}. 
Here it is assumed that the multi-TeV flaring in blazar occurs
within a compact and confined
region with a comoving radius $R'_f$ inside the blob of radius
$R'_b$\cite{Sahu:2013ixa} (henceforth $'$ implies jet comoving
frame).
In the context of leptonic model, the SED of the HBLs (the synchrotron
and the IC peaks) are fitted by taking into account different
parameters
(the blob radius $R'_b$, magnetic field $B'$, Doppler factor $\delta$,
bulk Lorentz factor $\Gamma$ etc.). For the present work, instead of discussing detail about
the SED of the individual HBLs, we use these best fit parameters from
the leptonic models which are shown in Table 1 and the references are
given for these objects.
As discussed earlier, in the inner region, the photon
density $n'_{\gamma,f}$ is very high compared to the photon density $n'_{\gamma}$ in
the outer region.
% i.e. $n'_{\gamma,f} \gg n'_{\gamma}$. 
The UHE protons undergo photohadronic interaction with the
seed photons in the inner region in the self-synchrotron Compton (SSC) regime through the
intermediate $\Delta$-resonance. In a normal blazar jet, however, the
photohadronic process is not the efficient mechanism to produce
multi-TeV  $\gamma$-rays and neutrinos because $n'_{\gamma}$ is low, 
which makes the optical depth $\tau_{p\gamma}\ll 1$. However, the assumption
of the compact inner jet region overcome this problem. The pion production in
$p\gamma$  collision through $\Delta$-resonance is 
\be
p+\gamma \rightarrow \Delta^+\rightarrow
 \left\{
\begin{array}{l l}
 p\,\pi^0, & \quad  {fraction~ 2/3}\\
  n\,\pi^+ ,
& \quad   {fraction~ 1/3}\\
\end{array} \right. .
\label{decaymode}
\ee
%\begin{array}{l l}
%p\,\pi^0, & \quad \text {fraction~ 2/3}\\ n\,\pi^+ , 
%& \quad  \text {fraction~ 1/3}\\
%\end{array} \right. ,
%\label{decaymode}
%\ee
%which has a cross section $\sigma_{\Delta}\sim 5\times 10^{-28}\,
%{\rm cm}^2$. 
The $\pi^+$ and $\pi^0$  will decay to GeV-TeV neutrinos and $\gamma$-rays respectively.
The optical depth of
the $\Delta$-resonance process in the inner compact region is 
$\tau_{p\gamma}=n'_{\gamma,f} \sigma_{\Delta} R'_f$, where $n'_{\gamma,f} $ is not known.
By assuming the Eddington luminosity is equally shared by the jet
and the counter jet in the blazar, for a given comoving photon energy
$\epsilon'_{\gamma}$ in the synchrotron/SSC regime we can get the
upper limit on the photon density as  
$n'_{\gamma,f} \ll L_{Edd}/(8\pi R'^2_f \epsilon'_{\gamma})$. We can
also compare the proton energy loss time scale
$t'_{p\gamma}\simeq (0.5\,n'_{\gamma,f}\sigma_{\Delta})^{-1}$ and the
dynamical time scale  $t'_{d}=R'_f/c$ in this region to estimate
$n'_{\gamma,f}$, so that 
the production of multi-TeV $\gamma$-rays and
neutrinos take place. Not to have over production of neutrinos and $\gamma$-rays, we
can assume a moderate efficiency (a few percents)  by taking 
$\tau_{p\gamma} < 1$ which gives $n'_{\gamma,f} < (\sigma_{\Delta}
R'_f)^{-1}$. The kinematical condition
for the production of $\Delta$-resonance in the observer's frame is 
\be
E_p \epsilon_{\gamma}=0.32 \frac{\Gamma {\delta}}{(1+z)^2} GeV^2,
\label{deltacondi}
\ee
where $E_p$ and $\epsilon_{\gamma}$ are the proton and the seed photon
energies respectively.
% The bulk Lorentz factor of jet is $\Gamma$ and
%the Doppler factor is given by $\delta$. 
In the decay of the $\Delta$-resonance to nucleon and pion, each pion carries $\sim 0.2$ of the
proton energy and from the pion decay each neutrino and photon carries
$1/4$ and $1/2$ of the pion energy respectively. So the individual neutrino
and photon energies are respectively $E_{\nu}=E_p/20$ and $E_{\gamma}=E_p/10$. This gives
\be
E_{\nu} \epsilon_{\gamma}=0.016 \frac{\Gamma {\delta}}{(1+z)^2} GeV^2.
%E_{\nu} \epsilon_{\gamma}=0.016\, \Gamma {\delta}{(1+z)^{-2}}\, { GeV}^2.
\label{neuph}
\ee
In a HBL, $\epsilon_{\gamma}$  can be calculated from the  given
neutrino energy if $\Gamma$ and $\delta$ are known. 
%The positional correlation of few IceCube events with HBLs, the success of
%photohadronic interaction to explain the multi-TeV emission from the
%flaring of some HBLs and the emission from Cen A opens an avenue to
%review the same hadronic model to give a possible explanation to these
%correlated IceCube events. For this reason, we use the hadronic model to
%estimate the $\epsilon_{\gamma}$ and the comoving photon density in
%the inner jet region to check the efficiency of the photohadronic process
%for objects having the positional correlation with the IceCube
%events. 
%We assume that in these blazars/AGN, the inner jet region is
%compact and dense compared to the outer region and not only the
%density of the SSC photons but also the low energy synchrotron photon
%density is high. 
%This high density photons can probably come from the
%accretion disk or produced within the inner jet region.
%\bwt

%%%%%%%%%%%%%%%%%%%%%%%%%%%%%%%%
\begin{table*}
\centering
\label{tab1}
\begin{tabular}{lcccccccc}
\hline
Object & ID & $\frac{E_{\nu}}{TeV}$ & $\frac{\epsilon_{\gamma}}{keV}$ &
$R'_{f,15}$ & $R'_{b,15}$& $n'_{\gamma,f,10}$ & $F_{\nu,-9}$ & {\bf $\delta\chi^2$} \\
%\vspace*{-0.2cm}
%{(Dec,RA);z,$\delta$}&\, & (TeV) & (keV) & (cm) & (cm) & (${\text cm}^{-3}$)
%                & (${\text erg}\,{\text cm}^{-2}{\text s}^{-1}$)\\
{(Dec,RA);z,$\delta$}&\, & \, &\,&\,&\,&\,&\,\\
%\vspace*{-0.2cm}
\hline
RGBJ0152+017\cite{Aharonian:2008xv} 
& 1& 47.6 & 179. &0.9& 1.5 & 2.2 &2.41 &  0.24\\
%\vspace*{-0.2cm}
(1.77,28.14);0.08,25&\,&\,&\,&\,&\,&\,&\,\\
%\hline
H2356-309\cite{Aharonian:2006zy} & 7 & 34.3 & 111. &0.5 & 3.4 & 4.0&2.38&  0.66\\
(-30.62,358.79); 0.165, 18 &10 & 97.2 & 39. & \, & \, & \,&\,&  0.47\\
                                  \,&21 & 30.2 & 125. & \, & \, &\,&\,&  0.29\\
%\hline
SHBLJ001355.9 \cite{Abramowski:2013daa} & 21 & 30.2 & 45. & 1.0 &35.&
2.0 & 2.41 & $0.13$\\        
( -18.89,3.46);0.095,10 &\,&\,&\,&\,&\,&\,&\,\\
%\hline
KUV00311-1938 &21&30.2&-&-&-&-&-&  0.05\\
(-19.35,8.39);-,-&\,&\,&\,&\,&\,&\,&\,\\
%\hline
Mrk421 \cite{Blazejowski:2005ih} & 9 & 63.2 & 46. & 3.0& 7.0 & 0.7&2.43& 0.61\\
(38.19,166.01); 0.031, 14 &\,&\,&\,&\,&\,&\,&\,\\
%\hline
1ES1011+496 \cite{Albert:2007kw} & 9 & 63.2 & 69. & 5.0&10. &0.4 & 2.36&0.94\\
(49.43,153.77);0.212,20 &\,&\,&\,&\,&\,&\,&\,\\
%\hline
PKS2005-489 \cite{Abramowski:2011as} & 12 & 104. & 31. & 5.0 & 400.&
0.4 &2.42&  0.33\\
(-48.83,302.36);0.071,15 &15&57.5&53.&\,&\,&\,& \,&  0.25\\
%\hline
PG1553+113 \cite{Aleksic:2011rq} & 17 & 200. & 50. & 3.0 &10. &0.7&2.29&  0.59\\
(11.19,238.94);0.4,35 & \,&\,&\,&\,&\,&\,&\,\\
%\hile
Mrk180 \cite{Costamante:2001ya} & 31& 42.5 & 34. & 5.0 &20. &0.4&2.43&  0.18\\
(70.16,174.11);0.045,10 & \,&\,&\,&\,&\,&\,&\,\\
1ES0502+675 \cite{Katarzynski:2011yc}& 31 & 42.5 & 35. & 5.0 &10. & 0.4&2.31&  0.66\\
(67.62,76.98);0.341,13 & \,&\,&\,&\,&\,&\,&\,\\
RGBJ0710+591 \cite{Acciari:2010qw} & 31 & 42.5 & 267.& 5.0 &20. &0.4 &2.39&   0.77\\
(59.15,107.61);0.125,30 & \,&\,&\,&\,&\,&\,&\,\\
1ES1312-423 \cite{Abramowski:2013pgm} & 35 & 2004. & 0.32 &5.0 & 240. &0.4 &2.40&  0.85\\
(-42.6,198.75);0.105,7. & \,&\,&\,&\,&\,&\,&\,\\
Cen A (FR-I) \cite{Sahu:2012wv} & 35 & 2004. & 0.056 &0 .6 & 3.0 &3.3 &2.45&  0.73\\
(-43.01,201.36);.00183,1 & \,&\,&\,&\,&\,&\,&\,\\

\hline
%\hline
\end{tabular}
\caption{The objects HBLs/AGN are  shown in first column which are in the error circles of
  the IceCube events ID (second column). Below each object we also put
  their coordinates, Declination and Right Ascension (Dec, RA) in degree,
  redshift (z) and the Doppler factor ($\delta$). In the third and the
  fourth columns the observed neutrino energy $E_{\nu}/TeV$  and 
the corresponding seed photon energy $\epsilon_{\gamma}/keV$ are given. In
fifth and the sixth columns the radius of the inner blob $R'_{f}$ and the outer blob $R'_{b}$
are given  in units of $R'=10^{15} R'_{15}\, cm$. The seed photon density
in the inner blob $n'_{\gamma,f}$ in units of 
$n'_{\gamma,f}=10^{10}\,n'_{\gamma,f,10}\,cm^{-3}$
  is given in the seventh column and diffuse neutrino flux $F_{\nu}$ in units of $F_{\nu}=10^{-9}\,
  F_{\nu,-9}\, GeV\,cm^{-2}\,s^{-1}\,sr^{-1}$ is given in the eighth
  column. In the last column we have shown the $\delta\chi^2$ value
  for each event. The reference to each object is given in the first column.}
\end{table*}
%\ewt
%%%%%%%%%%%%%%%%%%%%%%%%%%%%%%%%%%%%%%%%%%%%%%%

\section{Results}  
\label{sec:2}

We found  coincidence of the positions of 12 HBLs and
one radio galaxy, Cen A within the
error circles of 10 IceCube events. These HBLs and AGN  are taken from the
online catalog TeVCat\cite{TeVCat} and are observed in
multi-TeV $\gamma$-rays. However, the redshift, Lorentz
factor and doppler factor of some of these HBLs are not yet known. So 
whichever HBL has known $z$, $\Gamma$, $\delta$ and SED
and lies within the error circle of the IceCube event we calculate the
seed photon energy $\epsilon_{\gamma}$ necessary to produce the
desired neutrino energy $E_{\nu}$ through photohadronic interaction. 
The events 25 and 34 have very large errors $> 40^{\circ}$, so we
neglect these two events from our analysis.
For the calculation of $n'_{\gamma,f}$, first we estimate the
radius of the inner blob $R'_f$, which will satisfy the restriction
$R_s < R'_f < R'_b$, where $R_s=2 G_N M_{BH}/c^2$ is the Schwarzschild
radius of the  central black of mass $M_{BH}$. The 
$R'_b$ is obtained from the leptonic model fit to the  SED
of the object. The values of $R'_f$ and $R'_b$ for the objects
are shown in Table 1.
We assume  a very conservative $1\%$ energy loss of the UHE
protons in the inner blob on the dynamical time scale $t'_{d}$ which
corresponds to a optical depth of $\tau_{p\gamma}\sim 0.01$ and 
$n'_{\gamma,f}\sim 2\times 10^{10}\, R'^{-1}_{f,15}\, cm^{-3}$. The
proton in the inner jet region has maximum energy 
$E_{p,max}\sim 3\times 10^{17}(B'_f/G) R'_{f,15}\, eV$, where $B'_f$ is the comoving
magnetic field, which is higher than the
outer region. 
For all neutrino flavors $\alpha$  we assume a power-law spectrum of the form 
\be
J_{\nu_{\alpha}}(E_{\nu}) = A_{\nu_{\alpha} }\left (\frac{{E_{\nu}}}{{100\, TeV}}\right )^{-\kappa},
\ee
and the neutrino flux can be given as
\be
F_{\nu}=\sum_{\alpha} \int_{E_{\nu 1}(1+z)}^{E_{\nu 2}(1+z)}dE_{\nu} E_{\nu} J_{\nu_{\alpha}} (E_{\nu}).
\ee
The normalization factor
$A_{\nu_{\alpha}}$ is calculated by using the 988 days IceCube
data\cite{Aartsen:2013jdh}. The integration limit  
is from 25 TeV to 2.2 PeV\cite{Moharana:2015nxa} and $\kappa$
is the spectral index. For the luminosity distance calculation we take the
Hubble constant $H_0=69.6\,km\,s^{-1}\,Mpc^{-1}$,
$\Omega_{\Lambda}=0.714$ and $\Omega_m=0.286$.
All the 37 IceCube events with their individual error circles in 
equatorial coordinates are shown in the sky map in FIG. 1.
The 12 HBLs and the  Cen A are within the error circles of
10 IceCube events which are also shown in the sky map.
In Table 1,  we have summarized all the relevant parameters of these 13
objects. All the correlated IceCube events are shower events with sub-PeV energies and the event 
35 which is the only PeV event with $E_{\nu} \simeq 2$ PeV.  
Except the HBL, KUV00311-1993\cite{TeVCat}, all other have their $z$, $\Gamma$ and $\delta$
measured/fitted and SEDs are calculated
from the leptonic model.  For most of the objects
$\epsilon_{\gamma}$ lies between the synchrotron peak energy and the
forward falling tail of synchrotron energy with the exception of 
RGBJ0192+017\cite{Aharonian:2008xv} and
1ES1011+496\cite{Albert:2007kw}. In these two HBLs $\epsilon_{\gamma}$  lies in the beginning of
the SSC spectrum and the values  are 179 keV and 69 keV respectively. The corresponding photon densities 
and the neutrino fluxes are shown in Table 1.
%$n'_{\gamma,f}\sim 2.2\times 10^{10}\, {\text cm}^{-3}$ and $\sim
%4\times  10^{10}\, {\text cm}^{-3}$  and in the outer region are $n'_{\gamma}$  $\sim 3\times 10^3\, {\text cm}^{-3}$ and $\sim
%8\times  10^4\, {\text cm}^{-3}$ respectively. This shows that
%$n'_{\gamma,f}$ is many orders of magnitude larger than
%$n'_{\gamma}$. 
Our estimate of $n'_{\gamma,f}$  is based on the
assumption of $1\%$ energy loss of the UHECR proton for all the
HBLs/AGN. 
%The neutrino flux for these two HBLs are found to be
%$F_{\nu}\sim 2.41\times 10^{-9}\,GeV\,cm^{-2}\,s^{-1}\,sr^{-1}$ and $\sim
%2.36\times 10^{-9}\,GeV\,cm^{-2}\,s^{-1}\,sr^{-1}$ respectively 
We observed that
by varying $\kappa$  between 2.2 and 3.08
we found a small variation is the neutrino flux. So here
we  fix its value to 2.2.

The HBL, H2356-309\cite{Aharonian:2006zy} is within the error circles of three
IceCube events 7, 10 and 21 and their  corresponding synchrotron
energies, $n'_{\gamma, f}$ and neutrino flux are shown in Table 1. 
Another two HBLs, SHBLJ001355.9\cite{Abramowski:2013daa} and
KUV00311-1938 are also within the error circle of the event 21 and 
SHBLJ001355.9  has the corresponding synchrotron energy
$\epsilon_{\gamma}\simeq 45$ keV. 
%The synchrotron photon
%densities in the inner and the outer region of SHBLJ001355.9  are found to be $2\times
%10^{10}\,{\text cm}^{-3}$ and $\sim 4.2\times 10^3\,{\text cm}^{-3}$ respectively. 
The blazar PKS2005-489\cite{Abramowski:2011as} is in the error circles of the
events 12 and 15 and to produce these neutrino events  the photon
energy is in the range 30 keV-53 keV which is near the synchrotron
peak and the corresponding proton
energy is in the range $1.2\, {PeV} \le E_p \le 2.1\,{PeV}$. 
%The $1\%$ energy loss by the high energy proton
%corresponds to $n'_{\gamma,f}\sim 4\times 10^9\,{\text cm}^{-3}$ and
%the outer region has the seed photon density $n'_{\gamma} < 250\,{\text cm}^{-3}$. 
These two events are also
spatially correlated with the Fermi bubble. 
%and is possible that
%supernova remnant in this region can accelerate protons to high energy
%to produce neutrinos through $pp$ interaction. 
The event 17 has a mean energy of 200 TeV is
correlated with the HBL, PG1553+113\cite{Aleksic:2011rq} and is the farthest one in
our list with a redshift of $z=0.4$.
% The seed photon energy
%corresponding to this events is $\epsilon_\gamma\sim 50$ keV  and  the
%photon density in the inner region is $n'_{\gamma,f} \sim 7\times
%10^9\, {\text cm}^{-3}$ while the outer region has 
%$n'_{\gamma} \sim 7.4\times 10^6\, {\text cm}^{-3}$. Similarly, the
%neutrino flux is $\sim 2.29\times 10^{-9}\,
%GeV\,cm^{-2}\,s^{-1}\,sr^{-1}$.
The $n'_{\gamma, f}$ and neutrino fluxes for PKS2005-489 and PG1553+113 are shown in 
Table 1.

Very recently the Telescope Array (TA) observed an UHECR hotspot above 57 EeV in a region 
within $20^{\circ}$ radius circle centered at RA= $146.7^{\circ}$ and Dec. = $43.2^{\circ}$ \cite{Abbasi:2014lda}, 
the shaded closed counter in the sky map in FIG 1.
This region correlates with three neutrino events  9, 26 and
31. We found three HBLs: Mrk 180, 1ES0502+675 and RGBJ0710+591 within
the error circle of the IceCube event 31.
Interestingly, positions of two blazars,  Mrk 421\cite{Blazejowski:2005ih} and
1ES1011+496\cite{Albert:2007kw} are also simultaneously within the error circle of the
IceCube event  9 and within the  TA hotspot\cite{Padovani:2014bha,Fang:2014uja}. 
%The correlation of Mark 421 with both 
%IceCube events and TA hotspot is discussed very
%recently\cite{Padovani:2014bha, Fang:2014uja}.
The required $E_p$ and $\epsilon_{\gamma}$ for Mrk 421
are 1.3 PeV and 46 keV respectively. The photon density  and
$F_{\nu}$ are shown in Table 1. Similarly for 1ES1011+496 also we have shown the $n'_{\gamma,f}$ and
$F_{\nu}$ in Table 1.
%The  $\epsilon_{\gamma}\simeq 69$ keV corresponds to the photon
%density in the outer region $\sim 7.7\times 10^4\, {\text cm}^{-3}$. 
%For these two HBLs, the estimated neutrino fluxes are found to be
%$\sim 2.43\times 10^{-9}\, GeV\,cm^{-2}\,s^{-1}\,sr^{-1}$ and $\sim 2.36\times 10^{-9}\, GeV\,cm^{-2}\,s^{-1}\,sr^{-1}$ respectively.

%\bwt
%%%%%%%%%%%%%%%%%%%%%%%
\begin{figure}[t!]%fig1
%\vspace*{-1.6cm}
%%\vspace*{-0.7cm}
%%\hspace*{-.50cm}
%%\hspace*{-.70cm}
{\centering
%%\resizebox*{0.55\textwidth}{0.4\textheight}
\resizebox*{0.50\textwidth}{0.35\textheight}
{\includegraphics{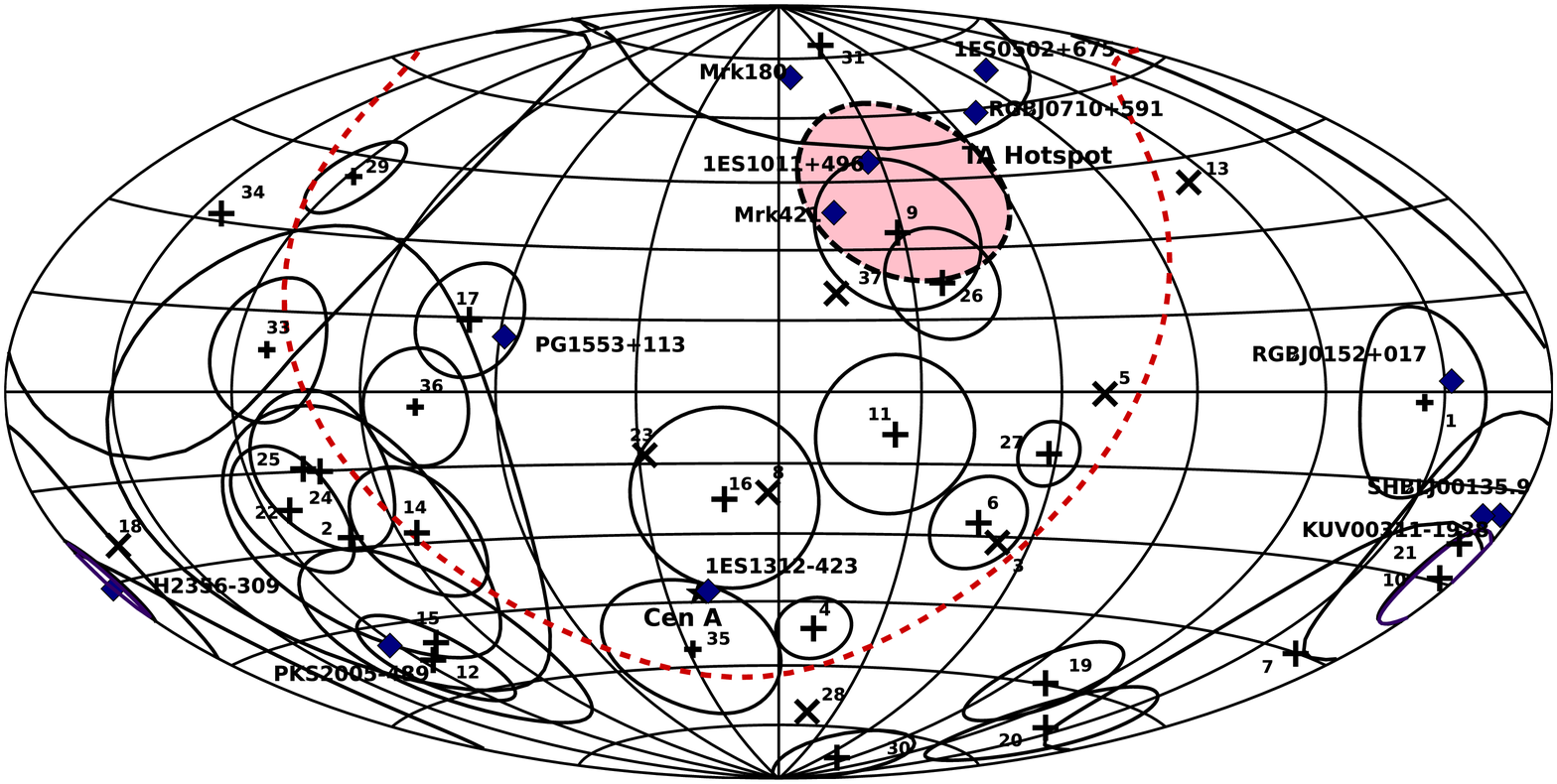}}
%{\includegraphics{Skymap.pdf}}
\par}
%\vspace*{-2.0cm}
{\centering
\caption{The sky map is shown in the equatorial coordinates with the 37
  IceCube events and their individual errors (only for shower
  events). Here $+$ are shower events and $\times$
  sign are track events with their corresponding event ID. We have also shown the
  positions of the HBLs with their names which are within the error
  circle of the  IceCube events.The TA hotspot
   is shown as a shaded closed contour and the galactic plane is
  shown as a dashed line. 
%(See the electronic edition of the Journal for a color version of this figure).
}
}
\label{eqskymap}
\end{figure}
%%%%%%%%%%%%%%%%%%%%%%%%%%%
%\ewt

Cen A is the nearest active  radio galaxy  and
long been proposed as the source of UHECRs.
Few years ago Pierre Auger (PA) Collaboration reported 
two UHECR events above 57 EeV within $3.1^{\rm o}$ around Cen
A\cite{Auger:2009}. 
Its position coincides within the error circle of the
IceCube event  35 having the highest neutrino energy of  2 PeV so far
observed by IceCube. 
In terms of the hadronic model discussed above
the 2 PeV neutrino energy corresponds to a proton energy of $\sim 40$ PeV and the seed photons energy is $\epsilon_{\gamma}\sim 56$
eV in the valley formed by the synchrotron and the SSC photons.
The seed photon density  $n'_{\gamma,f} \sim \times 10^{10}\, {cm}^{-3}$
around $\epsilon_{\gamma}\sim 56$ eV is also high.
%We obtain the neutrino flux 
%$F_{\nu}\sim 2.45\times 10^{-9}\, GeV\,cm^{-2}\,s^{-1}\,sr^{-1}$.
%For $\epsilon_{\gamma} \sim 56$ eV, the outer blob has seed photon
%density $n'_{\gamma} \sim 5.3\times 10^7\, {\text cm}^{-3}$. 
For $\epsilon_{\gamma}\, <\, 56$ eV, synchrotron emission dominates and
the low energy seed photon
density increases rapidly\cite{Sahu:2012wv}. So in principle
$E_{\nu} > 2$ PeV can be produced more
efficiently. But non-observation of neutrinos above 2 PeV from Cen A can be due
to (i) low flux of UHECR above 40 PeV and/or (ii) there is a cut-off
energy around 40 PeV beyond which the relativistic jet  is unable to
accelerate protons. Probably many more years of data are necessary to shed more
light on this possible correlation between the IceCube
event and the position of Cen A. 
Position of another HBL 1ES1312-423 also almost coincide with the
position of the Cen A and  thus falls within the error circle of the
IceCube event 35. For this HBL the $\epsilon_{\gamma}=0.32$ keV and
the corresponding observed photon flux is $F_{\gamma}\sim 6\times
10^{-12}\,erg\,cm^{-2}\,s^{-1}$ which is closed to the synchrotron peak\cite{Abramowski:2013pgm}.

The multi-TeV flaring of the objects 1ES1959+650, Mrk 421 and M87 are interpreted
through the photohadronic interaction as discussed in Section 2\cite{Sahu:2013cja,Sahu:2015tua,Sahu:2013ixa}. The
maximum energy of these high energy $\gamma$-rays are less than 20 TeV
(Mrk 421\cite{Sahu:2015tua}) which corresponds to proton energy $E_p <
200$ TeV and neutrino energy $E_{\nu}< 10$ TeV. But for
the interpretation of the IceCube events the necessary proton energy
will be $E_p=20\times E_{\nu}$. For $30\, {TeV} \le E_{\nu}\le 2\, {PeV}$
the proton energy will be in the range $600 \,{TeV} \le E_p \le 40\,
{PeV}$. So neutrino flux from the interaction of these very high energy protons
with the background photons can be small from an individual HBL.
Apart from this, we have only observed flaring episodes of very few
HBLs. So it is very hard to justify the temporal correlation of IceCube
events during a flaring episode of a HBL. We have to wait longer
period and have sufficient data to comment about the correlation
between the IceCube events and the flaring episode of the object.

%Due to the tiny neutrino mass then  can be time delay
%between the observed gamma rays events and the corresponding neutrinos
%events from these Mpc distance objects. Also apart from the
%correlation of Mrk 421 we have not seen position correlation between
%IceCube events and other flaring objects. It is difficult to draw
%any conclusion from only one correlation. So we have to wait for
%longer period to prove or disprove this hypothesis.

In the photohadronic scenario both the TeV-PeV neutrinos and
the TeV-PeV $\gamma$-rays are correlated as both are produced from the
decay of charged and neutral pions respectively as shown in
Eq.(\ref{decaymode}). The background seed photons responsible for the
production of these high energy neutrinos and $\gamma$-rays have energies
above few keV. These photons have energy in between the synchrotron
peak and the low energy tail of the SSC spectrum.
The TeV-PeV photons produced from the $\pi^0$ decay will interact
mostly with the same $\sim$ keV seed photons in the inner blob region to produce
$e^+e^-$ pairs. The required threshold energy for the seed photon to produce
the pair is $\epsilon_{\gamma,th}\ge 2 m_e^2/E_{\gamma}$ which is
mostly in the microwave range. Also the $\sigma_{\gamma\gamma}\sim
1.7\times 10^{-25}\, cm^{-2}$ is the maximum in the microwave
range and the pair creation cross section for keV  background
photon is very small $\sigma_{\gamma\gamma}\leq
10^{-29}\,cm^2$. In the region where the TeV-PeV photons and neutrinos
are produced, the microwave photon density is very low.
So even if the seed photon density is high (in the keV range), the mean 
free path for the TeV-PeV photons satisfy $\lambda_{\gamma\gamma}\gg R'_f$,
hence, there will be  negligible attenuation of these
photons in the inner blob region. Again in the outer blob,
the low energy photon density is order of magnitude smaller than 
the inner blob, so no attenuation in the outer region. 
However, on their way to the Earth, these TeV-PeV photons 
can interact with the low energy photons to produce pairs.

We have also done a statistical analysis to look for the correlation
between the IceCube events and the 42 TeV  emitting HBL/AGN from the
TeVCat\cite{TeVCat}. Here we
adopt the method used in ref.\cite{Moharana:2015nxa} and
convert the coordinates (RA and Dec) into unit vectors on a sphere as
\be
{\tilde x}=(sin\theta\, cos\phi, sin\theta\, sin\phi, cos\theta),
\ee
with $\phi=RA$ and $\theta=\pi/2-Dec$, 
where $i$ and $j$ correspond to the  event coordinates and object
coordinates respectively.
The angle between the two unit vectors ${\tilde x}_i$ and ${\tilde x}_j$ 
is given as
$\gamma=cos^{-1} ({\tilde x}_j.{\tilde x}_i)$ which is independent of the coordinate system and is a measure of correlation between the events and the objects.
Then one makes use of the quantity
\be
\delta\chi^2_i=min_j \left ( \gamma^2_{ij}/\delta\gamma^2_i\right )
\ee
where $\delta \gamma^2_i$ is the error on the
$i$th coordinate. Only 10 events meet the condition that $\delta\chi^2\leq 1$ with 13 objects which are shown in
the sky map and also in Table 1.  The $\delta\chi^2$ values of these
events are given in the last column of Table 1.
From the Monte Carlo simulation we estimate the significance of any
correlation with  IceCube events by randomizing the RA of the 42
  objects within their allowed ranges. One has to remember that the
  value of $\delta\chi$ for the object
closest to the neutrino event is chosen in this method.
The distribution of $\delta \chi^2_i$
is realized  by repeating this process one million times and the $p$-value is
calculated by counting the number of times 10 or more IceCube events
satisfy $\delta\chi^2\le 1$ divided by the total number of
realizations. In Fig. 2, the shaded histograms correspond to the
number of correlated neutrino events with the 42 objects
of the TeVCat in different ranges
of $\delta\chi^2$ value. The open histograms correspond to the
expected number of correlated neutrino events from the simulations
(continuous line for the randomized RA ) 
with their corresponding $p$-value which is $0.647$ corresponds to a
confidence level (CL) of $\sim 35\%$. 
In another simulation we select the IceCube events which have angular
errors $\le 20^{\circ}$. In this case the IceCube events 7, 21 and 31
will not contribute. So with this constraint in angular resolution, we
have only 7 events instead of 10 events considered earlier. In this
simulation we found the CL $\sim 42\%$ this is shown in Fig. 3.
Both of these analysis shows that there is no significant correlation
between the IceCube events and the HBLs positions. 
As we have shown by increasing the angular resolution from
$40^{\circ}$ to $20^{\circ}$ the CL increases by $\sim7\%$. Also we
believe that 42 objects from the TeVCat are not enough to give a
better statistics when the events are isotropic. Apart from these
objects there may be other type of sources which will contribute but are not included
in our list. In future we would like to consider more sources for our analysis.
%\bwt
%%%%%%%%%%%%%%%%%%%%%%%
\begin{figure}[t!]%fig2
%\vspace*{-0.2cm}
%%\vspace*{-0.7cm}
%%\hspace*{-.50cm}
%%\hspace*{-.70cm}
{\centering
%%\resizebox*{0.4\textwidth}{0.3\textheight}
\resizebox*{0.5\textwidth}{0.45\textheight}
%{\includegraphics[angle = -90]{statistics1.eps}}
{\includegraphics{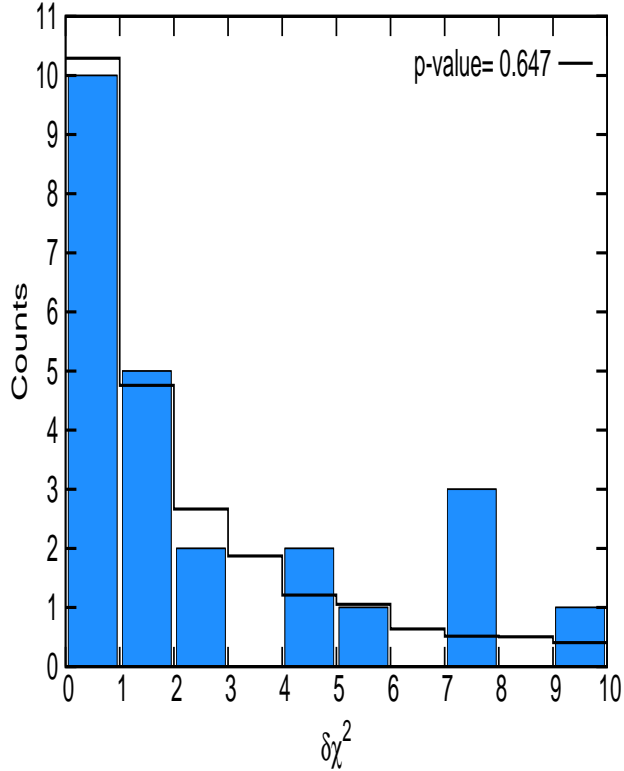}}
%{\includegraphics{statistics.pdf}}
\par}
\vspace*{-0.2cm}
{\centering
\caption{
The observed IceCube events (shaded histograms) and the simulated
events (open histograms with continuous line is for random RA) for different
$\delta\chi^2$ distribution are shown for angular resolution of the
IceCube events $\le40^{\circ}$. The $p$-value for the open
histograms are also given.
%(See the electronic edition of the Journal for a color version of this figure).
}
}
\label{stpro2}
\end{figure}
%%%%%%%%%%%%%%%%%%%%%%%%%%%
%\ewt

%\bwt
%%%%%%%%%%%%%%%%%%%%%%%
\begin{figure}[t!]%fig3
%\vspace*{-0.2cm}
%%\vspace*{-0.7cm}
%%\hspace*{-.50cm}
%%\hspace*{-.70cm}
{\centering
%%\resizebox*{0.4\textwidth}{0.3\textheight}
\resizebox*{0.5\textwidth}{0.45\textheight}
%{\includegraphics[angle = -90]{statistics2.eps}}
{\includegraphics{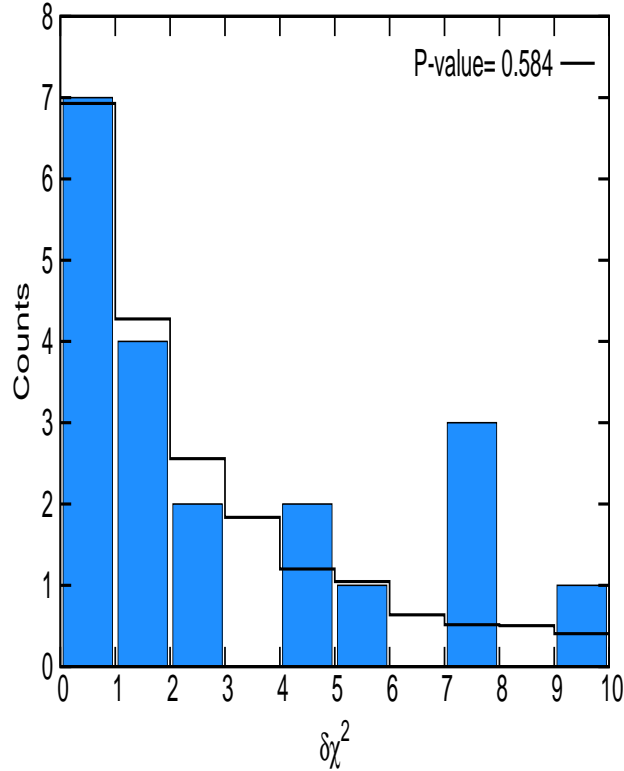}}
%{\includegraphics{statistics.pdf}}
\par}
\vspace*{-0.2cm}
{\centering
\caption{ Same as Fig. 2 but for angular resolution of the
IceCube events $\le 20^{\circ}$.
%The observed IceCube events (shaded histograms) and the simulated
%events (open histograms with continuous line is for random RA only and
%dotted line is for randomized Dec and RA) for different
%$\delta\chi^2$ distribution are shown. The $p$-values for the open
%histograms are also given.
%(See the electronic edition of the Journal for a color version of this figure).
}
}
\label{stat:3}
\end{figure}
%%%%%%%%%%%%%%%%%%%%%%%%%%%
%\ewt

\section{Conclusions} 
\label{sec:3}

The astrophysical interpretation of the 37 TeV-PeV neutrino events by
IceCube is challenging and several viable candidates have been
proposed and HBL is one of them.
The HBLs are the sources capable of producing multi-TeV
$\gamma$-rays. 
%The synchrotron and SSC photons are the dominant
%background in these objects. 
In the photohadronic scenario, TeV $\gamma$-rays are
accompanied with multi-TeV neutrinos from the decay of charged pions
and kaons. By analyzing  the online catalog TeVCat\cite{TeVCat}
we found coincidence of 12 HBLs and one FR-I galaxy Cen A positions within the
error circles of 10 IceCube events. All these events are
found to be shower events. The position of the HBL, H2356-309 coincides
with three IceCube events. We found positions of Mrk 421 and
1ES1011+496 are within the error circle of the IceCube event 9  as
well as within the error circle of the TA hotspot. The observed highest energy
PeV event  coincides with the positions of Cen A and the 
the HBL 1ES1312-423. Although, from the statistical analysis we found no significant correlation between
the IceCube events and the 42 objects in the TeV Catalog, it  does
not necessarily discard the photohadronic model interpretation for some of the IceCube
events. Many more years
of data are necessary to confirm or refute the positional
correlations of the HBLs/AGN with the IceCube events. 
Also these possible candidate sources should be constantly monitored and studied in greater detail to have
a better understanding of their properties and emission mechanisms.

%Many more years
%of data are necessary to confirm or refute the positional
%correlations of the HBLs/AGN with the IceCube events. 
%Also these possible candidate sources should be constantly monitored and studied in greater detail to have
%a better understanding of their properties and emission mechanisms.

We  thank Paolo Padovani, Kohta Murase, Alberto Rosales de Leon, S. Mohanty, S. Razzaque and Lukas Nellen
for many useful comments and discussions. The
work of S. S. is partially supported by DGAPA-UNAM (Mexico) Projects
No. IN103812 and IN110815.

\end{document}